\documentclass[pra,aps,a4paper,showpacs,twocolumn,floatfix]{revtex4-1}
\usepackage{graphicx}
\usepackage[ansinew]{inputenc}
\usepackage{array}
\usepackage{color}
\usepackage{amsmath}
\usepackage{amsxtra}
\usepackage{amstext}
\usepackage{amssymb}
\usepackage{latexsym}
\usepackage{dsfont}
\usepackage{verbatim}

\begin{document}

\title{Optically mediated nonlinear quantum optomechanics}

\author{H.~Seok$^1$, L.~F.~Buchmann$^1$, S.~Singh$^{1,2}$, and P.~Meystre$^1$}
\affiliation{$^1$ B2 Institute, Department of Physics and College of Optical
Sciences, University of Arizona, Tucson, AZ 85721\\
$^2$ ITAMP, Harvard-Smithsonian Center for Astrophysics, Cambridge, MA 02138, USA}

\begin{abstract}
We consider theoretically the optomechanical interaction of several mechanical modes with a single quantized cavity field mode for linear and quadratic coupling. We focus specifically on situations where the optical dissipation is the dominant source of damping, in which case the optical field can be adiabatically eliminated, resulting in effective multimode interactions between the mechanical modes. In the case of linear coupling, the coherent contribution to the interaction can be exploited e.g. in quantum state swapping protocols, while the incoherent part leads to significant modifications of cold damping or amplification from the single-mode situation. Quadratic coupling can result in a wealth of possible effective interactions including the analogs of second-harmonic generation and four-wave mixing in nonlinear optics, with specific forms depending sensitively on the sign of the coupling. The cavity-mediated mechanical interaction of two modes is investigated in two limiting cases, the resolved sideband and the Doppler regime. As an illustrative application of the formal analysis we discuss in some detail a two-mode system where a Bose-Einstein condensate is optomechanically linearly coupled to the moving end mirror of a Fabry-P\'erot cavity.  
\end{abstract}

\maketitle
\section{Introduction}

Quantum optomechanics has recently witnessed a rapid succession of key advances, with important milestones including the cooling of several systems to the quantum ground state of vibration~\cite{Cooling1, Cooling2, Cooling3, Cooling4}, the demonstration of strong optomechanical coupling~\cite{Cooling3, Strong1, Strong2} and of quantum coherent coupling between optical and phonon fields~\cite{Coherent}, the realization of optomechanical systems from ultracold atomic ensembles~\cite{BEC1, BEC2}, and impressive advances toward quantum state transfer~\cite{Transfer1, Transfer2, Transfer3, Transfer4, Transfer5, Transfer6}, to mention just a few examples. These developments hint at the promise for optomechanics applications in quantum metrology~\cite{Squeezing, PhononShotNoise}, quantum information science~\cite{Information1, Information2, Information3}, and also in possible tests of fundamental physics questions~\cite{Fundamental1, Fundamental2}. So far, though, the focus of most advances has been on single mode dynamics of the oscillator~\cite{Review1, Review2, Review3}, and only few authors have started to investigate the effects of multiple mechanical modes~\cite{multi1, multi2, multi3, Information3}. In complete analogy with the situation in quantum optics, however, single-mode quantum acoustics fails to capture many of its most important aspects, including most obviously perhaps propagation phenomena, nonlinear acoustical effects, focusing and defocusing of acoustic waves, etc. 

This paper represents a step toward the development of multimode quantum acoustics. The specific situation we consider is a multimode mechanical field -- perhaps provided by several mechanical oscillators or atomic samples trapped near extrema of an optical lattice, the mode coupling being provided by a common optical field. In some ways, this can be thought of as describing a situation analogous to collisions in atomic systems. Very much like collisions are effective interactions between atoms resulting from the elimination of the electromagnetic field, the optically mediated effective coupling between phononic modes can also be thought of in these terms. Generally speaking, collisions tend to be largely incoherent processes resulting e.g. in decoherence and dephasing in atomic ensembles, but it is known from ultracold atomic systems that this needs not be the case. In ultracold systems, such as quantum degenerate atomic gases, collisions can be a coherent  process, and have been demonstrated to lead to nonlinear atom optics effects~\cite{NLAO,Deng} such as e.g matter-wave four-wave mixing, soliton generation, the creation of entangled atom pairs, and more. In analogy to that situation, one can expect similar effects to occur in quantum acoustics. 

Our main goal is to develop a simple theory that shows that this is indeed the case, and that incoherent effects due to quantum fluctuations -- associated primarily with the elimination of the electromagnetic field -- need not overwhelm coherent effects. This paves the way to investigate nonlinear effects in phonon physics  reminiscent of those encountered in nonlinear optics and nonlinear atom optics; with the promise of similarly exciting applications.

\begin{figure}[]
\includegraphics[width=0.48 \textwidth]{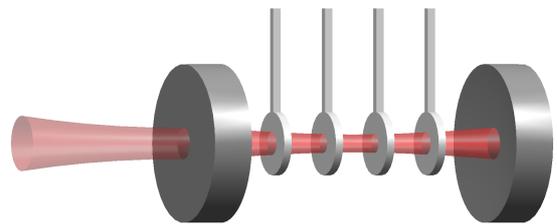}
\caption{\label{fig:system} Non-absorptive dielectric membranes inside a Fabry-P{\'e}rot resonator which interact with a single mode cavity field.}
\end{figure}

Our starting point is a relatively general discussion of optically mediated effective interactions between many mechanical modes, discussing in some detail the quantum noise and damping associated with that interaction, both for linear and quadratic optomechanical coupling. We identify a number of specific forms of the effective interaction that lead to qualitatively different dynamics, in particular a regime where the coupling of the quantum fluctuations of the mechanical modes is intrinsically nonlinear. We then illustrate the general analysis with an example involving just two mechanical modes, with emphasis on two limiting cases, the resolved sideband and the Doppler regimes. A specific example is given by the state transfer between two mechanical modes, for example a Bose-Einstein condensate coupled to the center-of-mass mode of a moving end mirror of a Fabry-P{\'e}rot resonator through a cavity field. We comment on the optimization of the  transfer fidelity using squeezed input light as the optically mediating light field and conclude with some general remarks and outlook.

\section{Model}

We consider a multimode optomechanical system described in terms of a set of $\cal{N}$ modes of effective masses $m_j$, bare frequencies $\omega_j$ and damping rates $\gamma_j$. The interaction between these modes is assumed to be mediated by a single mode of the optical resonator, of nominal frequency $\omega_c$ and driven by a monochromatic external field of frequency $\omega_L$. A realization of this system could be a high-finesse Fabry-P{\'e}rot resonator with a series of internal non-absorptive dielectric membranes such as depicted in Fig.~\ref{fig:system}, a collection of ultracold atomic samples trapped at or near the minima of an optical lattice inside such a resonator, one or several Bose-Einstein condensates trapped inside a Fabry-P\'erot cavity, perhaps with a moving end-mirror, or a number of similar setups operating either in the optical or in the microwave regime.

The Hamiltonian governing the system is
\begin{equation}
H = H_{\rm opt} + H_{\rm m} + H_{\rm om} +H_{\rm loss}, 
\end{equation} 
where 
\begin{equation}
H_{\rm opt} = \hbar\omega_c\tilde{A}^{\dag}\tilde{A}+i\hbar(\eta e^{-i\omega_Lt}\tilde{A}^{\dag}-\eta^{*} e^{i\omega_Lt}\tilde{A}),
\end{equation}
describes the cavity field mode, pumped externally by a field of frequency $\omega_L$ at a rate $\eta$, 
\begin{equation}
H_{\rm m} = \hbar\sum_{j=1}^{\cal N} \omega_{j}\hat{B}_{j}^{\dag}\hat{B}_{j}
\end{equation}
describes the mechanical modes, with $[\hat B_j, \hat B_k^\dagger] = \delta_{jk}$. The optomechanical interaction is 
\begin{equation}
H_{\rm om} = -\hbar\tilde{A}^{\dag}\tilde{A}\sum_{j=1}^{\cal N} g_{0,j}(\hat{B}_{j}+\hat{B}_{j}^{\dag})
\label{H linear}
\end{equation} 
in the linear case and
\begin{equation}
H_{\rm om} = \hbar\tilde{A}^{\dag}\tilde{A}\sum_{j=1}^{\cal N}g_{0,j}^{(2)}(\hat{B}_j+\hat{B}_j^{\dag})^2 
\label{H quadratic}
\end{equation}
in the quadratic case, with $g_{0,j}$ and $g_{0,j}^{(2)}$ linear and quadratic single-photon optomechanical coupling coefficients, respectively. Finally, $H_{\rm loss}$ describes the interaction of the cavity field and the mechanics with their respective reservoirs and accounts for dissipation. For simplicity, we will assume that all mechanical modes either couple purely linearly or quadratically. 

\section{Linear coupling}

This section discusses the case of linear coupling including modifications to cold damping and optical spring effects.

With 
\begin{equation}
\tilde{A}=\hat{A}e^{-i\omega_{L}t}
\end{equation}
and for the optomechanical coupling~(\ref{H linear}) the Heisenberg-Langevin equations of motion in the standard input-output formalism~\cite{InputOutput} are
\begin{eqnarray}
\dot{\hat{A}} &=&\left[i\Delta_c-\frac{\kappa}{2} \right] \hat{A}+\eta 
+ i\hat{A}\sum_{j=1}^{\cal N} g_{0,j}(\hat{B}_{j}+\hat{B}_{j}^{\dag})
+ \sqrt{\kappa} \hat {a}_{\rm in},  \nonumber \\
\dot{\hat{B}}_{j} &=& -i\omega_{j}\hat{B}_{j} -\frac{\gamma_{j}}{2}(\hat{B}_{j}-\hat{B}_{j}^{\dag}) +ig_{0,j}\hat{A}^{\dag}\hat{A} +i\tilde {\xi}_j,
\label{HL eq}
\end{eqnarray}
where
\begin{equation}
\Delta_c = \omega_{L}-\omega_{c},
\end{equation}
$\kappa$ and $\gamma_j$ are the decay rates of the cavity field and mechanical modes, respectively, and $\tilde{a}_{\rm in}$ and $\tilde{\xi}_j$ the associated Markovian quantum noise operators. Their two-time correlation functions will be given explicitly later on.

\subsection{Linearization}
If the system is driven by a classical field, it is useful to decompose the various field operators as the sum of their expectation values and small quantum fluctuations, whose dynamics are treated to lowest order only. Specifically we expand the operators $\hat A$ and $\hat B_j$ as
\begin{subequations} \label{mean field}
\begin{align}
\hat{A} &= \langle\hat{A}\rangle+\delta\hat{A} \equiv \alpha + \hat{a}, \\
\hat{B}_j &= \langle\hat{B}_j\rangle+\delta\hat{B}_j \equiv \beta_j + \tilde{b}_j,
\end{align}
\end{subequations}
where $[\hat a, \hat a^\dagger]=1$, $[\tilde b_j, \tilde b_k^\dag]=\delta_{jk}$ and we assume that their expectation values $\alpha$ and $\beta_j$ are much larger than the amplitudes of the fluctuations, for example, $|\alpha|^2 \gg \langle\hat{a}^{\dag}\hat{a}\rangle$. (Note that rigorously speaking this assumes all quantum contributions to have zero expectation value. However, the treatment remains valid for fluctuations with non-zero, but small expectation values~\cite{NZQuantum}.) This yields the mean field equations of motion
\begin{eqnarray}
\dot{\alpha} &=& i\bar{\Delta}\alpha-\frac{\kappa}{2}\alpha +\eta, \\
\dot{\beta}_j &=& -i\omega_{j}\beta_j-\frac{\gamma_{j}}{2}(\beta_j-\beta_j^{*}) +ig_{0,j}|\alpha|^2,
\end{eqnarray}
with steady-state values
\begin{eqnarray}
\alpha_{\rm s} &=& \frac{\eta}{-i\bar{\Delta}+\kappa/2} =\sqrt{\bar n_c},\\
(\beta_j+\beta^{*}_j)_{\rm s}&=&\frac{2g_{0,j}}{\omega_{j}}|\alpha_{\rm s}|^2,
\end{eqnarray}
where
\begin{equation}
\bar{\Delta} = \Delta_c +\sum_{j=1}^{\cal N} g_{0,j}(\beta_j+\beta_j^{*}).
\end{equation}
Here we have chosen the phase of $\eta$ such that $\alpha_s$ is real without loss of generality for the case of cw pumping.

As is well known, dynamical back-action gives rise to frequency shifts -- the optical spring effect -- and radiative damping or amplification for each mechanical oscillator~\cite{TheoryCooling1, TheoryCooling2, TheoryCooling3}. In order to capture these effects explicitly, we switch to an interaction picture with the transformations
\begin{eqnarray}
\tilde {b}_{j} &=& \hat{b}_{j}e^{-i\nu_jt},\\
\tilde {\xi}_{j} &=& \hat{\xi }_{j}e^{-i\nu_jt}
\end{eqnarray}
where 
\begin{equation}
\nu_j=\omega_j+\Omega_j,
\label{shifted freq}
\end{equation}
and the expressions for the dynamical shifts $\Omega_j$ will be determined self-consistently later on~\cite{SqueezingTransfer}. The equations of motion for the fluctuation operators then read
\begin{equation}
\dot{\hat{a}} = \left[i\bar{\Delta}-\frac{\kappa}{2}\right]\hat{a} +i\sum_{j=1}^{\cal N}g_{j}(\hat{b}_{j}e^{-i\nu_jt}+\hat{b}_{j}^{\dag}e^{i\nu_jt}) +\sqrt{\kappa}\hat{a}_{\rm in}, \label{deltadt} 
\end{equation}
\begin{equation}
\dot{\hat{b}}_j = \left[i\Omega_j-\frac{\gamma_{j}}{2}\right]\hat{b}_{j}+\frac{\gamma_{j}}{2}\hat{b}_{j}^{\dag}e^{2i\nu_jt} +ig_{j}e^{i\nu_jt}(\hat{a}^{\dag}+\hat{a}) +i\hat{\xi}_j, \label{deltabdt}
\end{equation}
where we have introduced the amplified optomechanical coupling strengths,
\begin{equation}
g_j = g_{0,j}\sqrt{\bar{n}_{\rm c}}.
\end{equation} 

\subsection{Elimination of the cavity field}

In the physically relevant regime where the optical loss is much larger than the mechanical decay rates, $\kappa \gg \gamma_j$, the cavity field follows the  dynamics of the mechanical oscillators adiabatically. For time scales long compared to $\kappa^{-1}$, the formal solution of Eq. (\ref{deltadt}) reads
\begin{eqnarray}
\hat{a}(t) &\simeq & -\sum_{j=1}^{\cal N}g_j\left[ \frac{e^{-i\nu_jt}\hat{b}_{j}}{(\bar{\Delta}+\nu_j)+i\kappa/2} +\frac{e^{i\nu_jt}\hat{b}_{j}^{\dag}}{(\bar{\Delta}-\nu_j)+i\kappa/2}\right] \nonumber \\ 
&&+\hat{f}_{\rm in}(t),
\end{eqnarray}
where the noise operator $\hat{f}_{\rm in}(t)$ is given by
\begin{equation}
\hat{f}_{\rm in}(t) = \sqrt{\kappa}\int_{0}^{t}d\tau ~e^{(i\bar{\Delta}-\kappa/2)(t-\tau)} \hat{a}_{\rm in}(\tau).
\end{equation}
Substituting the formal solution into the equations of motion for the mechanical oscillators, Eq. (\ref{deltabdt}), gives 
\begin{eqnarray}
\dot{\hat{b}}_j&=& \left[i\Omega_j-\frac{\gamma_{j}}{2}\right]\hat{b}_{j}+\frac{\gamma_{j}}{2}\hat{b}_{j}^{\dag}e^{2i\nu_jt} \nonumber \\
&-& \sum_{k=1}^{\cal N}g_{jk}\left[i\Omega_{k}+\frac{\Gamma_{k}}{2}\right]e^{i(\nu_j-\nu_k)t}\hat{b}_{k} \nonumber \\
&-& \sum_{k=1}^{\cal N}g_{jk}\left[i\Omega_{k}-\frac{\Gamma_{k}}{2}\right]e^{i(\nu_j+\nu_k)t}\hat{b}_{k}^{\dag} \nonumber \\
&+& ig_{j}e^{i\nu_jt}(\hat{f}_{\rm in}+\hat{f}_{\rm in}^{\dag}) +i \hat{\xi}_j 
\label{eombeforeRWA},
\end{eqnarray} 
where
\begin{equation}
g_{jk} = g_j/g_k.
\end{equation}
and the radiation-induced frequency shifts $\Omega_k$ and damping coefficients $\Gamma_k$ are given by
\begin{subequations}\label{shiftsdamping}
\begin{align}
\Omega_{k} &= g_{k}^2\left[\frac{\bar{\Delta}-\nu_k}{(\bar{\Delta}-\nu_k)^2+\kappa^2/4} +\frac{\bar{\Delta}+\nu_k}{(\bar{\Delta}+\nu_k)^2+\kappa^2/4}
\right], \label{omegaj}
\end{align}
\begin{align}
\frac{\Gamma_{k}}{2} &=  g_{k}^2\left[\frac{\kappa/2}{(\bar{\Delta}+\nu_k)^2+\kappa^2/4} -\frac{\kappa/2}{(\bar{\Delta}-\nu_k)^2+\kappa^2/4}
\right].\label{gammaj}
\end{align}
\end{subequations}

Due to the appearance of the shifted frequencies $\nu_k(\bar \Delta)$ in the denominators on its right hand side, Eq.~(\ref{omegaj}) is a fifth-order polynomial equation for $\Omega_k(\bar \Delta)$, so that it is not Lorentzian in general. In the strong coupling regime $g_k > \kappa/2$, it becomes multivalued for a range of effective detunings, a feature that can be interpreted in terms of normal mode splitting~\cite{Strong1}. In this situation the mechanical and optical modes are strongly hybridized and it is not meaningful to treat them separately. Also, the counter-rotating terms in Eqs.~(\ref{eombeforeRWA}), combined in particular with a driving field blue detuned from the resonator resonance, can result in instabilities in the strong coupling regime and a concomitant would lead to the break-down of the linearization process. 

The strong coupling and blue detuned regimes will be the object of future work. Here we focus on the simpler case of weak coupling, $g_k \ll \kappa$. In this limit the  frequency shifts $\Omega_k$ remain small compared to the bare frequencies of the oscillators, so that we can approximate  $\nu_k$ by $ \omega_k$ on the right hand side of Eqs.~(\ref{shiftsdamping}), see Eq.~(\ref{shifted freq}). The frequency shifts $\Omega_k$ are then the sum of two dispersion curves while the decay rates $\Gamma_k$ consist of two Lorentzians with opposite signs, all of which are centered around $\bar\Delta=\pm\omega_k$. The weak coupling regime also allows us to drop the counter-rotating terms in Eqs.~(\ref{eombeforeRWA}).
  
The resulting equations of motion are
\begin{eqnarray}
\dot{\hat{b}}_j &\simeq & \left[i\Omega_j-\frac{\gamma_j}{2}\right]\hat{b}_{j} -\sum_{k=1}^{\cal N}g_{jk}\left[i\Omega_{k}+\frac{\Gamma_{k}}{2}\right]e^{i(\nu_j-\nu_k)t}\hat{b}_{k} \nonumber \\
&+&ig_{j}e^{i\nu_jt}(\hat{f}_{\rm in}+\hat{f}_{\rm in}^{\dag}) +i \hat{\xi}_j, \label{RWA}
\label{weakcoupling linear}
\end{eqnarray} 
These equations describe a system of ${\cal N}$ pairwise, linearly coupled oscillators, with coherent contributions proportional to the optically induced frequency shifts $\Omega_k$, cold damping coefficients  $\Gamma_k$ familiar from the single-mode situation, and associated noise operators $(\hat{f}_\mathrm{in}+\hat{f}_\mathrm{in}^{\dag})$ and $\hat{\xi}_j$. The fact that each mechanical element experiences a different frequency shift and radiative decay, depending on its optomechanical coupling strength and the pump-cavity detuning, is the key ingredient that allows one to couple oscillators with different bare frequencies on-demand, by choosing appropriate pump-powers and detunings.  We discuss selected consequences of this coupled dynamics in section V for the case of a two-mode system. But first, the next section extends our discussion to the case of quadratic optomechanical coupling.

\section{Quadratic  coupling}
We now consider the quadratic optomechanical coupling
\begin{equation}
H_{\rm om} = \hbar\tilde{A}^{\dag}\tilde{A}\sum_{j=1}^{\cal N}g_{0,j}^{(2)}(\hat{B}_j+\hat{B}_j^{\dag})^2, 
\end{equation}
see Eq.~(\ref{H quadratic}), where $g_{0,j}^{(2)}$ are single-photon coupling coefficients. In that case the Heisenberg-Langevin equations of motion are, in the frame of the laser frequency $\omega_L$,
\begin{align}
\dot{\hat{A}} &= \left[i\Delta_{\rm c}-\frac{\kappa}{2}\right]\hat{A}+\eta -i\hat{A}\sum_{j=1}^{\cal N}g_{0,j}^{(2)}(\hat{B}_j+\hat{B}_j^{\dag})^2 + \sqrt{\kappa}\hat{a}_{\rm in}, \nonumber \\
\dot{\hat{B}}_j &= -i\omega_j\hat{B}_j-\frac{\gamma_j}{2}(\hat{B}_j-\hat{B}_j^{\dag}) -2ig_{0,j}^{(2)}\hat{A}^{\dag}\hat{A}(\hat{B}_j+\hat{B}_j^{\dag}) + i\tilde{\xi}_j \nonumber
\end{align}
with $\tilde A = \hat A \exp(-i\omega_L t)$ as before. With the expansion~(\ref{mean field}) we have 
\begin{align}
\dot{\alpha} &\simeq  i\bar{\Delta}^{(2)}\alpha -\frac{\kappa}{2}\alpha +\eta, \nonumber \\
\dot{\beta}_j &\simeq  -i\omega_j\beta_j-\frac{\gamma_j}{2}(\beta_j-\beta_j^{*})-2ig_{0,j}^{(2)}|\alpha|^2(\beta_j+\beta_j^{*}),
\end{align}
where the effective detuning is now
\begin{equation}
\bar{\Delta}^{(2)} \simeq  \Delta_{\rm c} -\sum_{j=1}^{\cal N}g_{0,j}^{(2)}(\beta_j + \beta_j^{*})^2,
\label{shifted delta}
\end{equation}
and consistently with the linearization procedure we have neglected in the second line the vacuum expectation value contribution $\langle \hat b \hat b^\dagger\rangle=1$. The steady-state expectation value of the cavity field amplitude is
\begin{equation}
\alpha_{\rm s} = \frac{\eta}{-i\bar{\Delta}^{(2)}+\kappa/2},
\end{equation}
and the steady-state displacements of the mechanical oscillators are found from the ${\cal N}$ coupled fifth-order equations
\begin{equation}
\left(4g_{0,j}^{(2)}|\alpha_{\rm s}|^2 +\omega_j\right)(\beta_j+\beta_j^{*})_{\rm s} =0.
\label{ssquad}
\end{equation} 
Zero displacement $(\beta_j+\beta_j^{*})_{\rm s}=0$ is always a solution to these equations, and for $g_{0,j}^{(2)} > 0$, which is the case for membranes located at minima of the intracavity field, it is the only stable solution. However that solution may become unstable for negative couplings, $g_{0,j}^{(2)}<0$, occurring when oscillators are located at cavity-field maxima. In this case, each oscillator settles in one of the stable  positions, 
\begin{equation}
\bar x_j = \pm \frac12 (\beta_j+\beta_j^*)_{\rm s},
\end{equation}
located symmetrically around the intracavity field maximum.  
 As we will see in the following sections, these two geometries result in qualitative  consequences for the behavior of the system. 
\subsection{Local maxima}

To simplify the discussion we assume that all coupling constants $g^{(2)}_{0,j}$ have the same sign, and consider first the case where they are negative. The stable steady state is degenerate, as each oscillator can have a negative or positive mean displacement. A particular choice of steady state will  break the symmetry of the sign of the displacements and influence the physics of the system. Experimentally, it could be controlled by addressing the individual oscillators through an additional field or preparing them slightly displaced towards the desired steady-state position. 
The Heisenberg-Langevin equations of motion for the fluctuations can be approximated to lowest non-vanishing order, as
\begin{eqnarray}
\dot{\hat{a}} &\simeq & \left[i\Delta^{(2)}  -\frac{\kappa}{2}\right]\hat{a}-4i\sum_{j=1}^{\cal N}g_j^{(2)}\bar{x}_j(\hat{b}_j+\hat{b}_j^{\dag}) \nonumber \\
&+& \sqrt{\kappa}\hat{a}_{\rm in},\\
\dot{\hat{b}}_j &\simeq & -\left[i\varpi_j+\frac{\gamma_j}{2}\right]\hat{b}_j -\left[2ig_{0,j}^{(2)}\bar{n}_{\rm c}-\frac{\gamma_j}{2}\right]\hat{b}_j^{\dag} \nonumber \\
&-& 4ig_j^{(2)}\bar{x}_j(\hat{a}+\hat{a}^{\dag})+i\tilde{\xi}_j, 
\end{eqnarray}
where we have introduced the amplified quadratic optomechanical strengths
\begin{equation}
g_j^{(2)} = g_{0,j}^{(2)}\sqrt{\bar{n}_{\rm c}} 
\end{equation}
and the shifted detuning of the cavity field and mechanical frequencies are
\begin{eqnarray}
\Delta^{(2)} &=& \bar{\Delta}^{(2)}-4\sum_{j=1}^{\cal N}g_{0,j}^{(2)}\bar{x}_{j}^2=\Delta_c-8\sum_{j=1}^{\cal N} g_{0,j}^{(2)}  \bar x_j^2, \\
\varpi_j &=& \omega_j+2g_{0,j}^{(2)}\bar{n}_{\rm c}.
\label{cor delta}
\end{eqnarray}
Note that in contrast with the situation for linear optomechanical coupling the correction $8\sum_{j=1}^{\cal N} g_{0,j}^{(2)}  \bar x_j^2$ to the detuning $\Delta_c$ is now twice as large as the correction to the mean-filed dynamics, compare Eq.~(\ref{cor delta}) to Eq.~(\ref{shifted delta}). This second-harmonic generation process, a direct acoustic analog of optical second harmonic generation,  is a direct consequence of the nonlinear nature of the quadratic optomechanical coupling and is driven by the mean-field mechanical oscillations.

Assuming that the resulting frequencies of the mechanical oscillators are
$$\nu_j=\varpi_j+\Omega_j,$$ 
and following the same approach as before the equations of motion for the mechanical modes become
\begin{eqnarray}
\dot{\hat{b}}_j &=& \left[i\Omega_j-\frac{\gamma_j}{2}\right]\hat{b}_j -\left[2ig_{0,j}^{(2)}\bar{n}_{\rm c}-\frac{\gamma_j}{2}\right]\hat{b}_j^{\dag}e^{2i\nu_jt} \nonumber \\
&-&\sum_{k=1}^{\cal N}g_{jk}^{(2)}\left[i\Omega_{k}+\frac{\Gamma_{k}}{2}\right]e^{i(\nu_j-\nu_k)t}\hat{b}_{k} \nonumber \\
&-&\sum_{k=1}^{\cal N}g_{jk}^{(2)}\left[i\Omega_{k}-\frac{\Gamma_{k}}{2}\right]e^{i(\nu_j+\nu_k)t}\hat{b}_{k}^{\dag} \nonumber \\
&-&4ig_{j}^{(2)}\bar{x}_je^{i\nu_jt}(\hat{f}_{\rm in}+\hat{f}_{\rm in}^{\dag}) +i\hat{\xi}_j, \label{localmaximabeforeRWA}
\end{eqnarray} 
where the noise operator associated with the adiabatically eliminated optical field is 
\begin{equation}
\hat{f}_{\rm in}(t) = \sqrt{\kappa}\int_{0}^{t}d\tau ~e^{(i\Delta^{(2)}-\kappa/2)(t-\tau)} \hat{a}_{\rm in}(\tau)
\end{equation}
and
\begin{equation}
g_{jk}^{(2)} = \frac{g_{j}^{(2)}\bar{x}_j}{g_{k}^{(2)}\bar{x}_k}. 
\label{quad coupling 1}
\end{equation}
We then find for the radiation-induced frequency shifts $\Omega_{k}$ and damping coefficients $\Gamma_{k}$ 
\begin{align}
\Omega_{k} &= \left(4g_{k}^{(2)}\bar{x}_k\right)^2\nonumber\\
&\times \left[\frac{\Delta^{(2)}-\nu_k}{(\Delta^{(2)}-\nu_k)^2+\kappa^2/4} +\frac{\Delta^{(2)}+\nu_k}{(\Delta^{(2)}+\nu_k)^2+\kappa^2/4} \right],\\
\frac{\Gamma_{k}}{2} &= \left(4g_{k}^{(2)}\bar{x}_k\right)^2 \nonumber\\
&\times \left[\frac{\kappa/2}{(\Delta^{(2)}+\nu_k)^2+\kappa^2/4} -\frac{\kappa/2}{(\Delta^{(2)}-\nu_k)^2+\kappa^2/4}  \right].
\end{align}

In the weak coupling regime, $g_j^{(2)}\bar{x}_j \ll \kappa$, we can neglect the fast oscillating terms on the right-hand side of Eq.~(\ref{localmaximabeforeRWA}) so that the equations of motion for the mechanics can be further simplified to
\begin{eqnarray}
\dot{\hat{b}}_j &=& \left[i\Omega_j-\frac{\gamma_j}{2}\right]\hat{b}_j -\sum_{k=1}^{\cal N}g_{jk}^{(2)}\left[i\Omega_{k}+\frac{\Gamma_{k}}{2}\right]e^{i(\nu_j-\nu_k)t}\hat{b}_{k} \nonumber \\
&&-4ig_{j}^{(2)}\bar{x}_je^{i\nu_jt}(\hat{f}_{\rm in}+\hat{f}_{\rm in}^{\dag}) +i \hat{\xi}_j. 
\end{eqnarray} 

These equations have the same form as in the case of linear optomechanical coupling of the mechanical oscillators, the main new features in this regime are second-harmonic generation and the flexibility to choose the sign and strength of the coupling coefficients $g_{jk}^{(2)}$ through the equilibrium positions $\bar x_j$ and $\bar x_k$, see Eq.~(\ref{quad coupling 1}).

\subsection{Local minima}

The situation is changed qualitatively in the case where the mechanical oscillators are located at minima of the cavity field. The single-photon, single mode quadratic optomechanical coupling coefficients are then positive and the cavity field results in a tighter confinement through the static optical spring effect. The only stable steady-state displacement is given by  $\bar{x}_{j}=0$ so that the effects described in the previous section are highly suppressed. On the other hand, a number of novel features appear as a result of the fact that  the first non-vanishing couplings are of higher order in the quantum fluctuations. In particular, instead of the second-harmonic generation discussed in the previous section, the dominant nonlinear effects are akin to four-wave mixing in optics. 

To lowest order in the fluctuations, the Heisenberg-Langevin equations of motion become
\begin{eqnarray}
\dot{\hat{a}} &\simeq & \left[i\bar{\Delta}^{(2)}-\frac{\kappa}{2}\right]\hat{a} -i\sum_{j=1}^{\cal N}g_{j}^{(2)}(\hat{b}_j^2+\hat{b}_j^{\dag 2}+2\hat{b}_j^{\dag}\hat{b}_j)
 \nonumber \\
&&+\sqrt{\kappa}\hat{a}_{\rm in},\\
\dot{\hat{b}}_j &\simeq & -\left[i\varpi_j+\frac{\gamma_j}{2}\right]\hat{b}_j -\left[2ig_{0,j}^{(2)}\bar{n}_{\rm c}-\frac{\gamma_j}{2}\right]\hat{b}_j^{\dag} \nonumber \\
&&-2ig_{j}^{(2)}(\hat{a}+\hat{a}^{\dag})(\hat{b}_j+\hat{b}_j^{\dag})+i\tilde{\xi}_j, 
\end{eqnarray}
where the effective detuning $\bar{\Delta}^{(2)} $ is now 
\begin{equation}
\bar{\Delta}^{(2)} = \Delta_{\rm c} -\sum_{j=1}^{\cal N}g_{0,j}^{(2)}.
\end{equation}
Applying the same arguments and approximations as before to adiabatically eliminate the optical field dynamics, we find the coupled equations of motion for the mechanical modes
\begin{widetext}
\begin{eqnarray}
\dot{\hat{b}}_j &=& \left[i\Omega_j-\frac{\gamma_j}{2}\right]\hat{b}_j -\left[2ig_{0,j}^{(2)}\bar{n}_{\rm c}-\frac{\gamma_j}{2}\right]\hat{b}_j^{\dag}e^{2i\nu_jt} -\frac{1}{2}\sum_{k=1}^{\cal N}g_{jk}^{(2)} \left[i\Omega_k+\frac{\Gamma_k}{2}\right](e^{-2i\nu_kt}\hat{b}_k^{2}\hat{b}_j +e^{2i(\nu_j-\nu_k)t}\hat{b}_k^{2}\hat{b}_j^{\dag}) \nonumber \\
&-&\frac{1}{2}\sum_{k=1}^{\cal N}g_{jk}^{(2)} \left[i\Omega_k-\frac{\Gamma_k}{2}\right](e^{2i\nu_kt}\hat{b}_k^{\dag2}\hat{b}_j +e^{2i(\nu_j+\nu_k)t}\hat{b}_k^{\dag2}\hat{b}_j^{\dag}) -\frac{i}{2}\sum_{k=1}^{\cal N}g_{jk}^{(2)}\Lambda_k(\hat{b}_k^{\dag}\hat{b}_k\hat{b}_j +\hat{b}_k^{\dag}\hat{b}_k\hat{b}_j^{\dag}e^{2i\nu_jt}) \nonumber \\
&-&2ig_{j}^{(2)}(\hat{f}_{\rm in}+\hat{f}_{\rm in}^{\dag})(\hat{b}_j+\hat{b}_j^{\dag}e^{2i\nu_jt}) +i\hat{\xi}_j,
\label{bdot minima}
\end{eqnarray}
\end{widetext}
where the mode-coupling is now given by
\begin{equation}
g_{jk}^{(2)} = g_{j}^{(2)}/g_{k}^{(2)},
\end{equation}
the frequency shifts and radiative damping coefficients are
\begin{eqnarray}
\Omega_{k} &=& (2g_{k}^{(2)} )^2\\
&\times& \left[\frac{\bar{\Delta}^{(2)}-2\nu_k}{(\bar{\Delta}^{(2)}-2\nu_k)^2+\kappa^2/4} +\frac{\bar{\Delta}^{(2)}+2\nu_k}{(\bar{\Delta}^{(2)}+2\nu_k)^2+\kappa^2/4} \right], \nonumber \\
\frac{\Gamma_{k}}{2} &=&  (2g_{k}^{(2)} )^2\\
&\times& \left[\frac{\kappa/2}{(\bar{\Delta}^{(2)}+2\nu_k)^2+\kappa^2/4} -\frac{\kappa/2}{(\bar{\Delta}^{(2)}-2\nu_k)^2+\kappa^2/4}
\right],\nonumber
\end{eqnarray}
\begin{equation}
\Lambda_k=(2g_{k}^{(2)} )^2 \frac{2 \bar \Delta^{(2)}}{(\bar \Delta^{(2)})^2 + \kappa^2/4},
\end{equation}
and 
\begin{equation}
\hat{f}_{\rm in}(t) = \sqrt{\kappa}\int_{0}^{t}d\tau e^{(i\bar{\Delta}^{(2)}-\kappa/2)(t-\tau)} \hat{a}_{\rm in}(\tau).
\end{equation}
Note that in Eq. (\ref{bdot minima}) the quantum noise stemming from the elimination of the optical field is now multiplicative.

Since we can neglect fast rotating terms in the weak coupling regime, Eqs.~(\ref{bdot minima}) further simplify  to
\begin{eqnarray}
\dot{\hat{b}}_j &=& \left[i\Omega_j-\frac{\gamma_j}{2}\right]\hat{b}_j -\frac{i}{2}\sum_{k=1}^{\cal N}g_{0,jk}^{(2)}\Lambda_k\hat{b}_k^{\dag}\hat{b}_k\hat{b}_j\nonumber \\
&-&\frac{1}{2}\sum_{k=1}^{\cal N}g_{0,jk}^{(2)} \left[i\Omega_k+\frac{\Gamma_k}{2}\right]e^{2i(\nu_j-\nu_k)t}\hat{b}_k^{2}\hat{b}_j^{\dag} \nonumber \\
&-&2ig_{j}^{(2)}(\hat{f}_{\rm in}+\hat{f}_{\rm in}^{\dag})(\hat{b}_j+\hat{b}_j^{\dag}e^{2i\nu_jt}) +i\hat{\xi}_j. \label{localminimaRWA}
\end{eqnarray}
These coupled nonlinear equations are indicative of four-wave mixing processes, and are formally similar to situations driven by $\chi^{(3)}$ nonlinear susceptibilities in nonlinear classical and quantum optics ~\cite{Nonlinear, Quantum}. In that context, four-wave mixing phenomena are known to lead to a rich spectrum of effects that have found numerous applications in applied and fundamental studies. They include for example sum frequency generation, high harmonic generation, optical parametric amplification, self-focusing and defocusing, the generation and propagation of optical solitons, phase conjugation, the generation of quantum mechanical squeezing and entangled photon pairs, and more~\cite{Nonlinear, Quantum}. The analysis of this section paves the way to investigate a similar range of studies in quantum optomechanics and quantum acoustics, and as such, opens the way to an intriguing and rich direction of investigation that will be carried out in follow-up studies. One particularly attractive feature of optomechanics in this context is the ease with which the sign and strength of the nonlinear interactions can be adjusted in combination with the functionalization of these devices. The following sections discuss a few illustrative examples, concentrating for now on the simple situation of a two-mode system with linear optomechanical coupling only, and demonstrating a example of functionalization that involves the optically mediated coupling of  a nanomechanical system and a quantum degenerate atomic system. 

\section{Example -- two-mode system}
Elementary multimode effects can be studied in the case where two mechanical modes interact linearly with a common cavity field, i.e. ${\cal N}=2$. In the weak coupling regime the Heisenberg-Langevin Eqs.~(\ref{weakcoupling linear}) become
\begin{eqnarray} 
\dot{\hat{b}}_1 &=& -\frac12\Gamma_{e,1}\hat{b}_1 -\left[i\Omega_{c,1}+\frac12\Gamma_{c,1}\right]\hat{b}_2 e^{i(\nu_1-\nu_2)t} \nonumber \\
&+&ig_2e^{i\nu_1 t}(\hat{f}_{\rm in}+\hat{f}_{\rm in}^{\dag}) +i \hat{\xi}_1\nonumber \\
\dot{\hat{b}}_2 &=& -\frac12 \Gamma_{e,2}\hat{b}_2 -\left[i\Omega_{c,2}+\frac12\Gamma_{c,2}\right]\hat{b}_1e^{-i(\nu_1-\nu_2)t} \nonumber \\
&+&ig_2e^{i\nu_2t}(\hat{f}_{\rm in}+\hat{f}_{\rm in}^{\dag}) +i \hat{\xi}_2
\label{2modes}
\end{eqnarray}
where we have introduced the effective decay rate
\begin{equation}
\Gamma_{e,j} = \Gamma_j+\gamma_j,
\end{equation}
the cross-damping rate
\begin{equation}
\Gamma_{c,j} = g_{jk}\Gamma_k,
\label{cross damping}
\end{equation}
and the cross-coupling frequency
\begin{equation}
\Omega_{{c},j} = g_{jk}\Omega_k.
\label{cross coupling}
\end{equation}
The cross-damping rate is associated with effects such as sympathetic cooling or heating of coupled oscillators, while the cross-coupling frequency is associated with the coherent aspects of multimode coupling, such as e.g. quantum state transfer. Note that when the two effective mechanical frequencies are  matched, $\nu_1=\nu_2$, the cross-coupling frequencies and damping coefficients become equal $\Omega_{{c},1}=\Omega_{{c},2},~\Gamma_{{c},1}=\Gamma_{{c},2}$. 

In terms of the motional quadratures
\begin{eqnarray}
\hat{X}_j &=& \frac{1}{2}(\hat{b}_je^{-i\nu_jt}+\hat{b}_j^{\dag}e^{i\nu_jt}), \\
\hat{P}_j &=& \frac{1}{2i}(\hat{b}_je^{-i\nu_jt}-\hat{b}_j^{\dag}e^{i\nu_jt}),
\end{eqnarray}
Eqs.~(\ref{2modes}) become
\begin{equation}
\dot{\hat{u}} = \mathcal{M}\hat{u} +\hat{\Xi},
\end{equation}
where $\hat{u} =  \begin{pmatrix}
  \hat{X}_1, 
  \hat{P}_1, 
  \hat{X}_2, 
  \hat{P}_2
 \end{pmatrix}^{T}$,
 the drift matrix $\mathcal{M}$ is given by
 \begin{eqnarray}
  \mathcal{M} = \frac12 \begin{pmatrix}
    -\Gamma_{e,1} & 2\nu_1 & -\Gamma_{c,1}& 2\Omega_{c,1}\\
  -2\nu_1 & -\Gamma_{e,1} & -2\Omega_{c,1} & -\Gamma_{c,1}\\
  -\Gamma_{c,2}& 2\Omega_{c,2} & -\Gamma_{e,2} & 2\nu_2\\
  -2\Omega_{c,2} & -\Gamma_{c,2}& -2\nu_2 & -\Gamma_{e,2}
 \end{pmatrix},
 \end{eqnarray}
and the noise operator matrix $\hat{\Xi}$ is
\begin{eqnarray}
\hat{\Xi} = 
 \begin{pmatrix}
  0 \\
  \hat{\xi}_1e^{-i\nu_1 t} +g_1\hat{F}_{\rm in}  \\
  0 \\
  \hat{\xi}_2e^{-i\nu_2 t}+g_2\hat{F}_{\rm in} 
 \end{pmatrix},
\end{eqnarray}
with the effective optical noise operator
\begin{equation}
\hat{F}_{\rm in} = \sqrt{\kappa}\int_{0}^{t}d\tau ~e^{(i\bar{\Delta}-\kappa/2)(t-\tau)} \hat{a}_{\rm in}(\tau)+{\rm adj.}.
\end{equation}
In the following we limit ourselves to the case of Gaussian states and assume that the mechanical elements are coupled to statistically independent Markovian heat baths at low temperature $T$ characterized by the two-time correlations functions~\cite{mechanicalnoise}
\begin{align}
\frac{\langle\tilde{\xi}_j(t)\tilde{\xi}_j(t')+\tilde{\xi}_j(t')\tilde{\xi}_j(t)\rangle}{2} &\simeq  \gamma_j(2\bar{n}_{{\rm th},j}+1)\delta(t-t^{\prime}),
\end{align}
where $\bar{n}_{{\rm th},j} = [\exp(\hbar\omega_j/k_BT)-1]^{-1}$. From Eqs.~(\ref{mean field}) the first moments of all position and momentum quadratures are always zero, and one readily derives a closed set of differential equations for the second moments of the oscillator quadratures. The equation of motion for the covariance matrix $V_{ij}=\frac{1}{2}\langle\hat{u}_i\hat{u}_j+\hat{u}_j\hat{u}_i\rangle$ can be written as {\cite{Mari}}
\begin{equation}
\dot{V} = \mathcal{M}V+V\mathcal{M}^{T}+D,
\end{equation}
where $\cal{M}$ is the drift matrix, and the inhomogeneous term $D$ is given by $D_{ij}=\frac{1}{2}\langle\hat{\Xi}_{i}\hat{u}_{j}+\hat{u}_{j}\hat{\Xi}_{i} +\hat{\Xi}_{j}\hat{u}_{i}+\hat{u}_{i}\hat{\Xi}_{j}\rangle$. This system of equations can be solved exactly in principle but the resulting solutions lack transparency. More physical insight into the coupled dynamics of the oscillator can be gained in the resolved sideband limit and in  the Doppler limit, two cases to which we now turn.

\subsection{Resolved sideband regime}

\begin{figure}[]
\includegraphics[width=0.48 \textwidth]{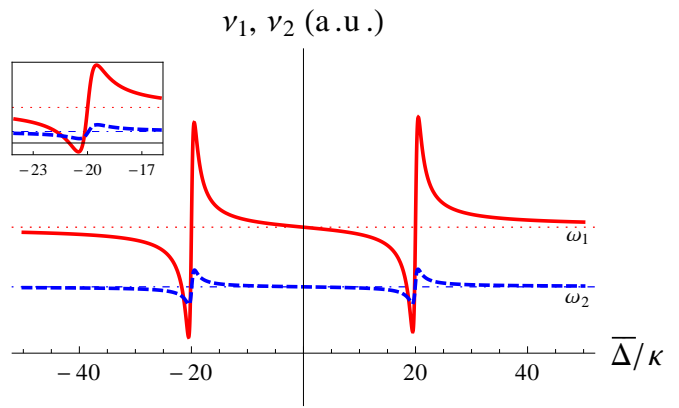}
\caption{\label{fig:frequency_matching} Effective frequencies of the first (red, solid) and second (blue, dashed) mechanical modes in the resolved sideband regime $\omega_1, ~\omega_2 \gg \kappa$ for $g_1 > g_2$. The dotted lines denote the bare frequencies of the mechanical modes. The shifted frequencies are matched at the intersections. The inset gives a detailed view on the intersections occurring in the red-detuned side. The parameters used are $\omega_1=2\pi\times20$MHz, $\omega_2=2\pi\times19.95$MHz, $\kappa=2\pi\times1$MHz, $g_1=2\pi\times0.3$MHz, $g_2=2\pi\times0.12$MHz.}
\end{figure}

The resolved sideband regime is characterized by slow optical resonator damping compared to the mechanical frequencies $\omega_1,~\omega_2 \gg \kappa$. (We assume without loss of generality that $\omega_1 > \omega_2$ in the following.) It is known from the standard theory of single-mode cavity optomechanics that in the absence of coupling each oscillator experiences a frequency shift  $\Omega_k(\bar \Delta)$ comprised of the sum of two dispersion curves centered around $\bar{\Delta}=\pm\omega_k$ with characteristic widths $\kappa$, see Fig.~\ref{fig:frequency_matching}, while their damping coefficients $\Gamma_k(\bar \Delta)$ are the sum of a positive and negative Lorentzian, also of width $\kappa$, and centered at these same detunings~\cite{Review4}.  

The red and blue detunings for which the two oscillators are brought into resonance with each other are given by the intersections of the two curves in Fig.~\ref{fig:frequency_matching}. There are two such intersections on the red-detuned side of the optical resonance, one near $\bar \Delta = -\omega_k$ and one on the wing of the dispersion curve. Two more intersections are located similarly on the blue-detuned side. At these points we have $\nu_1 = \nu_2 \equiv \nu $ and therefore $\Omega_{c,1} = \Omega_{c,2} \equiv \Omega_c$ and $\Gamma_{c,1} = \Gamma_{c,2} \equiv \Gamma_c$ from Eq.~(\ref{cross coupling}). It is for those detunings that the coupling between the two mechanical oscillators is most effective. 

\begin{figure}[]
\includegraphics[width=0.48 \textwidth]{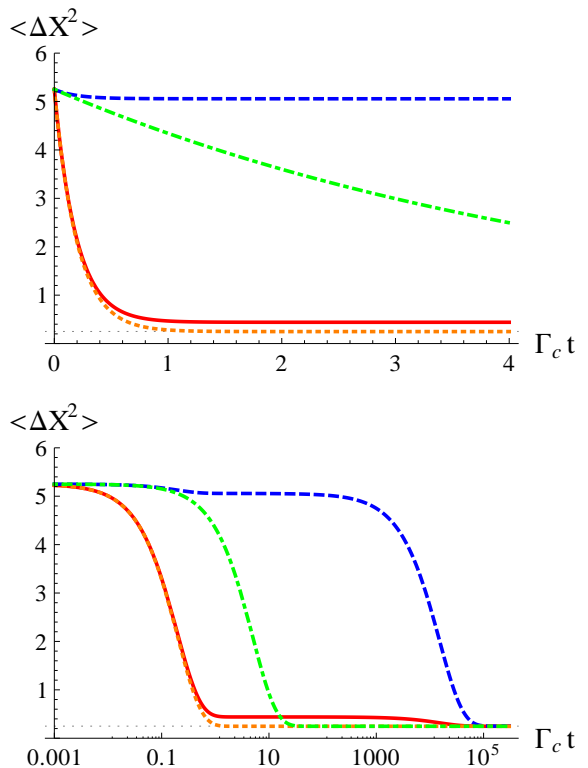}
\caption{\label{fig:heating} Upper plot: short  time (linear time scale) and lower plot: long time (log time scale) evolution of the motional quadratures of the mechanical modes in the single and two-mode scenarios. The mechanical modes of frequencies $\omega_1$ (red, solid) and $\omega_2$ (blue, dashed) are both cooled down while interacting with each other. For comparison the uncoupled cold damping of the modes $\omega_1$ (orange, dotted) and  $\omega_2$ (green, dotdashed) are also shown. Two-mode coupling  results in slowing down of the individual cold damping rates.  Here $\omega_1=2\pi\times20$MHz, $\omega_2=2\pi\times19.99$MHz, $\kappa=2\pi\times0.95$MHz, $g_1=2\pi\times50$kHz, $g_2=2\pi\times10$kHz.}
\end{figure}

The detailed dynamics of resonant mode coupling depends on whether the resonance condition $\nu_1 = \nu_2$ occurs near the center or on the wing of the dispersion curves that comprise  $\Omega_k(\bar \Delta)$. Near the resonance $\bar \Delta = \pm \omega_k$ optical damping (or amplification) dominates over the optical spring effect, see Eqs.~(\ref{omegaj}) and (\ref{gammaj}), and to first approximation one can safely neglect that effect compared to those of $\Gamma_{c,k}$ by setting $\Omega_{c,k}=0$, see Eqs.~(\ref{cross damping}) and (\ref{cross coupling}). The drift matrix ${\mathcal M}$ reduces to
 \begin{eqnarray}
  \mathcal{M} = \frac{1}{2}
  \begin{pmatrix}
    -\Gamma_{e,1} & 2\nu & -\Gamma_{c} & 0\\
  -2\nu & -\Gamma_{e,1} & 0 & -\Gamma_{c}\\
  -\Gamma_{c} & 0 & -\Gamma_{e,2} & 2\nu\\
  0 & -\Gamma_{c} & -2\nu & -\Gamma_{e,2}
 \end{pmatrix}
\label{driftmatrix}
 \end{eqnarray}
and the dominant consequence of the effective optomechanical coupling between the oscillators is the modification of their rate of cold damping or amplification compared to their uncoupled values. One such example is shown in Fig.~\ref{fig:heating}, which plots the evolution of the variances of the motional quadratures of the mechanical modes when their effective frequencies are matched around the center of the Lorentzian curve on the red side of the cavity field. The mechanical modes are both assumed to be initially in thermal equilibrium. In the absence of cross interaction, both oscillators are subject to cold damping (dotted and dot-dashed lines) with cooling rates that differ since $g_1 \neq g_2$. The coupling changes the situation significantly, with both cooling rates now markedly slower (solid and dashed lines). This is a result of the exchange of thermal excitations between the two oscillators that inhibits their direct optical cooling. (Note that our numerical calculations suggest that the long-time limit and thus the cooling limit does not change.) Performing a normal-mode analysis on the two modes reveals that the two out-of-phase oscillations decouple from the cavity field, while the two in-phase modes are cooled with an increased damping rate. 

We now turn to the resonances $\nu_1 = \nu_2$ that occur on the wings of the dispersion curves. In these cases the optical spring effect dominates over cold damping (or amplification), a consequence of the faster decay of the Lorentzian $\Gamma_k(\bar \Delta)$ (with $1/\bar \Delta^2$) compared to the dispersive nature of $\Omega_k(\bar \Delta)$, which decreases as $1/\bar \Delta$. That situation is therefore characterized by the dominance of coherent exchange between the two oscillators, and we approximate the corresponding drift matrix by
 \begin{eqnarray} \label{transfer_matrix}
  \cal{M} = \begin{pmatrix}
    -\gamma_1/2 & \nu & 0 & \Omega_c\\
  -\nu & -\gamma_1/2 & -\Omega_c & 0\\
  0 & \Omega_{{c}} & -\gamma_2/2 & \nu\\
  -\Omega_c & 0 & -\nu & -\gamma_2/2
 \end{pmatrix}.
 \end{eqnarray}
In this regime, and provided that the noise sources associated with the optical field can be managed, one can realize coherent effects such as quantum state transfer between the two mechanical modes. We return to this effect in the Doppler regime, which we consider in the following.

\subsection{Doppler regime}
\begin{figure}[]
\includegraphics[width=0.48 \textwidth]{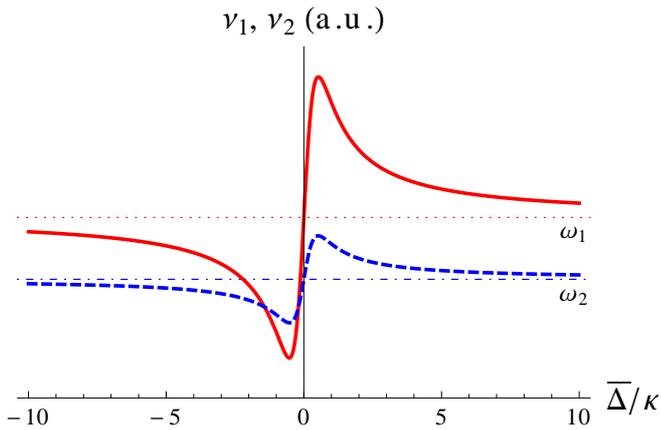}
\caption{\label{fig:Doppler_frequency_matching} Qualitative behavior of the effective mechanical frequencies for the first (red, solid) and second (blue, dashed) oscillators in the Doppler regime. The shifted frequencies are matched at the intersections. Here we chose $g_1>g_2$. A set of parameters is $\omega_1=2\pi\times100$kHz, $\omega_2=2\pi\times93$kHz, $\kappa=2\pi\times1$MHz, $g_1=2\pi\times90$kHz, $g_2=2\pi\times50$kHz.}
\end{figure}

In the Doppler regime, $\omega_1,~\omega_2 \ll \kappa$, the width of the optical cavity is so large that it washes out the two distinct features of $\Omega_k(\bar \Delta)$ and $\Gamma_k(\bar \Delta)$, characteristic of the resolved sideband regime, and one can approximate
\begin{eqnarray}
\Omega_j &\simeq & g_j^2\frac{2\bar{\Delta}}{\bar{\Delta}^2+\kappa^2/4},\label{dopplershifts} \\
\Gamma_j &\simeq & -g_j^2\frac{2\bar{\Delta}\kappa \nu_j}{(\bar{\Delta}^2+\kappa^2/4)^2}.
\end{eqnarray}
The qualitative behavior of the effective frequencies $\nu_j \,(j= 1,2)$ is plotted in Fig. ~\ref{fig:Doppler_frequency_matching} for $g_1>g_2$ and shows the intersections of the two curves on the red-detuned side of the cavity resonance. (They would lie on the blue-detuned side for $g_2>g_1$). Since $\Gamma_j(\bar \Delta)$ decays with the third power of the cavity linewidth while $\Omega_j(\bar \Delta)$ scales as $1/\kappa^2$, it follows that for the large cavity damping rates, characteristic of the Doppler regime, we can neglect in first approximation the effects of cold damping (or amplification) and at resonance we find again
 \begin{eqnarray}
  \cal{M} = \begin{pmatrix}
  -\gamma_1/2 & \nu & 0 & \Omega_c\\
  -\nu & -\gamma_1/2 & -\Omega_c & 0\\
  0 & \Omega_c & -\gamma_2/2 & \nu\\
  -\Omega_c & 0 & -\nu & -\gamma_2/2
 \end{pmatrix}.
 \end{eqnarray}
 Note that the argument leading to this result is different in the resolved sideband limit, where the radiation induced damping can be neglected if the effective detuning is outside of the mechanical sideband, and in the Doppler regime, in which case it is negligible due to the large cavity linewidth.

\subsection{State transfer}
The previous discussion showed that both, the resolved sideband and the Doppler regimes, provide situations where cold damping can be neglected and the two oscillators are resonantly coupled by the effective coherent interaction 
\begin{equation}
H_{\rm e} = \hbar\Omega_{c}(\hat{b}_1\hat{b}_2^{\dag}+\hat{b}_1^{\dag}\hat{b}_2).
\end{equation}
This is the familiar beam splitter Hamiltonian, which is known in particular to lead to quantum state swapping between the two oscillators. However, an important issue is that the optical field mediating the interaction also results in the appearance of quantum noise, further limiting the fidelity of state transfer in addition  to the clamping noise associated with the mechanical coupling to  thermal reservoirs. Clamping noise can be reduced in principle in cryogenic environments or perhaps in levitated structures~\cite{Levitation}. What we show in this brief section is that the fundamental quantum noise associated with the optical field can also be reduced in principle by the use of squeezed input fields with correlation functions~\cite{Gardiner}

\begin{eqnarray}
\langle\hat{a}_{\rm in}(t)\hat{a}_{\rm in}(t^{\prime})\rangle &=& e^{-i\omega_L(t+t^{\prime})}M\delta(t-t^{\prime}), \\
\langle\hat{a}^{\dag}_{\rm in}(t)\hat{a}^{\dag}_{\rm in}(t^{\prime})\rangle &=& e^{i\omega_L(t+t^{\prime})}M^*\delta(t-t^{\prime}), \\
\langle\hat{a}^{\dag}_{\rm in}(t)\hat{a}_{\rm in}(t^{\prime})\rangle &=& N\delta(t-t^{\prime}), \\
\langle\hat{a}_{\rm in}(t)\hat{a}^{\dag}_{\rm in}(t^{\prime})\rangle &=& (N+1)\delta(t-t^{\prime}),
\end{eqnarray}
where $\omega_L$ is the central frequency of a squeezing device, and positive-valued $N$ and complex-valued $M$ are the squeezing parameters which define the strength of the squeezing as well as its direction in phase space, given by the complex phase of $M$. In the following, we assume an ideal squeezed state which satisfies
\begin{equation}
M = \sqrt{ N(N+1)}e^{-2i\theta_s}.
\end{equation}
The equations of motion for vacuum input noise are recovered by substituting $N \rightarrow 0,~M \rightarrow 0$.

For concreteness we consider the explicit situation of a hybrid two-mode mechanical system where one of the modes is a recoil-induced side-mode of a Bose-Einstein condensate~\cite{BEC1} trapped inside an optical resonator with an oscillating end-mirror, and the other is the center-of-mass mode of oscillation of  that mirror. As discussed in Ref.~\cite{Transfer5}, in the case where a system is prepared in such a way that the shifted optomechanical frequencies of the two oscillators $\nu_1$ and $\nu_2$ are equal, $\Omega_{c,1} = \Omega_{c,2} = \Omega_c$, the equations of motion for the quadratures of the coupled mechanical oscillators reduce to 
\[
 \frac{d}{dt}
\begin{bmatrix}
\hat X_1  \\
\hat P_1  \\
\hat X_2  \\
\hat P_2
\end{bmatrix}
=
\begin{bmatrix}
0 & \nu & 0 & \Omega_c \\
-\nu & 0 & -\Omega_c & 0 \\
0 & \Omega_c& 0 & \nu \\
-\Omega_c & 0 & -\nu & 0
 \end{bmatrix}
\begin{bmatrix}
\hat X_1  \\
\hat P_1  \\
\hat X_2  \\
\hat P_2
\end{bmatrix}
+
 \begin{bmatrix}
0  \\
g_1\hat{F}_{\rm in}  \\
0  \\
g_2\hat{F}_{\rm in}
\end{bmatrix},
\]
where the subscripts 1 and 2 refer to the oscillating mirror and the condensate side-mode respectively, and the cross-coupling frequency is
\begin{equation}
\Omega_c=g_1g_2\frac{2 \bar \Delta}{\bar \Delta^2+\kappa^2/4}.
\end{equation}
For a squeezed input field the two-time correlation function of the effective noise can be approximated as
\begin{eqnarray}
&&\frac{\langle\hat{F}_{\rm in}(t)\hat{F}_{\rm in}(t')+\hat{F}_{\rm in}(t')\hat{F}_{\rm in}(t)\rangle}{2} \approx \frac{\kappa}{\bar{\Delta}^2+\kappa^2/4}\times \nonumber \\
&&\left(|M|\cos[2(\theta_s-\theta_c)]+N+\frac{1}{2}\right)\delta(t-t'),
\end{eqnarray}
with $\theta_c = \arg(i\bar{\Delta}+\kappa/2)$. Here we have assumed that intrinsic mechanical decoherence is negligible for the time scales of interest.

\begin{figure}[]
\includegraphics[width=0.48 \textwidth]{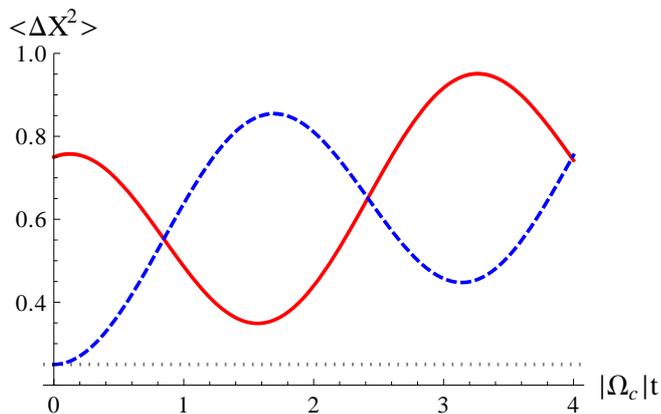}
\caption{\label{fig:thermal_BEC} Motional quadrature of the moving mirror (red, solid) and the BEC (blue, dashed) in the presence of vacuum noise for  $\omega_1=2\pi\times101$kHz, $\omega_2=2\pi\times100$kHz, $\kappa=2\pi\times1$MHz, $g_1=2\pi\times100$kHz, $g_2=2\pi\times10$kHz.}
\end{figure}

Figure~\ref{fig:thermal_BEC} shows the co-evolution of the variance of one of the motional quadratures of the two oscillators. In this example the resonator is driven by a coherent field and the initial states of the BEC and the moving mirror are the ground state and a thermal state with mean phonon number $\bar{n}=1$, respectively. As expected, the fidelity of the coherent quantum state transfer, which would be unity at times $(\pi/2 + p \pi)\Omega_c^{-1}$, with $p$ integer, in the absence of noise, is increasingly reduced by the quantum fluctuations of the light field, even without intrinsic mechanical decoherence mechanisms. 
\begin{figure}[]
\includegraphics[width=0.48 \textwidth]{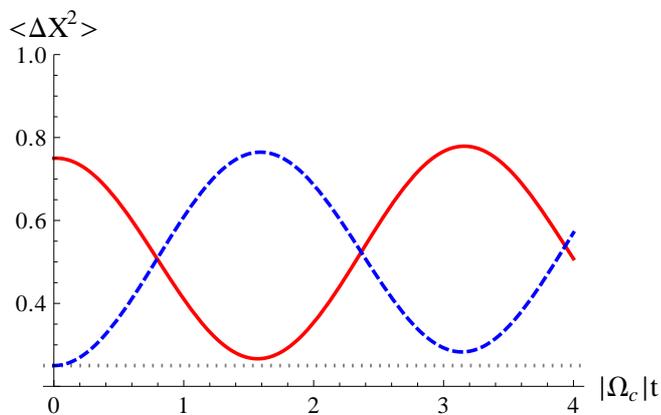}
\caption{\label{fig:squeezed_BEC} Motional quadrature of the moving mirror (red, solid) and the BEC (blue, dashed) for squeezed vacuum noise with $N=1$ and squeezing phase $\theta_s-\theta_c=\pi/2$. Same parameters as in Fig~\ref{fig:thermal_BEC}.}
\end{figure}

\begin{figure}[]
\includegraphics[width=0.48 \textwidth]{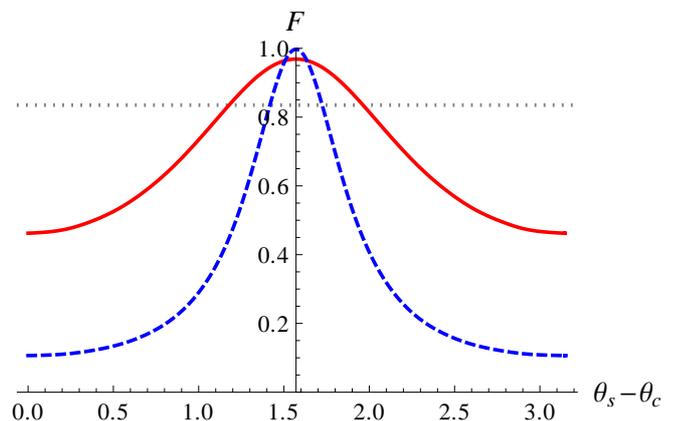}
\caption{\label{fig:Fidelity_BEC} State transfer fidelity for squeezed vacuum noise with $N=1$ (red, solid),  $N=10$ (blue, dashed) and vacuum noise (gray, dotted) as a function of the phase difference between the squeezing input and the cavity field. The initial states of the BEC and the moving mirror are the ground state and a thermal state with mean phonon number $\bar{n}=1$, respectively.}
\end{figure}

\begin{figure}[]
\includegraphics[width=0.48 \textwidth]{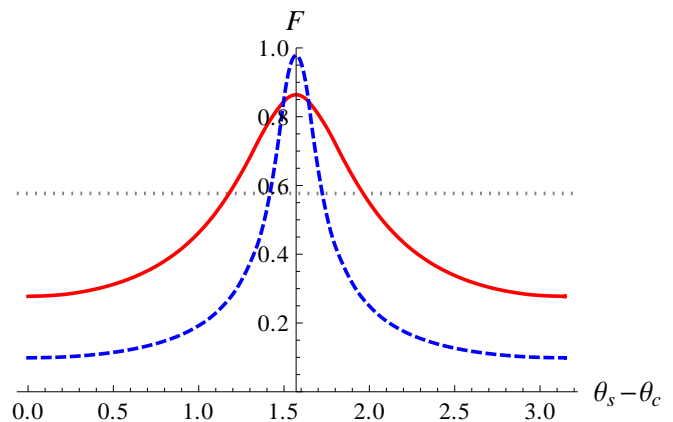}
\caption{\label{fig:Fidelity_BEC_squeezed} Same as Fig.~\ref{fig:Fidelity_BEC}, but for the BEC initially in a squeezed state with position quadrature variance of $0.025$ and momentum quadrature variance of $10$ and and the moving mirror in a thermal state with mean phonon number $\bar{n}=1$, respectively. }
\end{figure}

The use of a squeezed optical field to reduce the detrimental effect of quantum noise is illustrated in Fig.~\ref{fig:squeezed_BEC}, where the improved fidelity is readily apparent in particular at time $t=\pi/2\Omega_c$.  As illustrated in Figs.~\ref{fig:Fidelity_BEC} and ~\ref{fig:Fidelity_BEC_squeezed}, the state transfer fidelity depends strongly on the phase difference between the cavity field and a squeezing device. For $\theta_s =\theta_c\pm\pi/2$, the fluctuations in the position quadrature of the cavity field are minimized so that very high fidelity state transfer is achieved, $\mathcal{F}=0.997$ for $N=10$ in our example. As would be intuitively expected, we also find that strong squeezing leads to an increased sensitivity of the state transfer fidelity on that relative phase. As a comparison to Fig.~\ref{fig:Fidelity_BEC}, which is for a BEC side-mode initially in the ground state and the oscillating mirror in a thermal state with $\bar n = 1$, Fig.~\ref{fig:Fidelity_BEC_squeezed} displays the fidelity of state transfer where the BEC is initially in a squeezed state with $\langle\Delta\hat{X}_2^2\rangle=0.025$ and $\langle\Delta\hat{P}_2^2\rangle=10$. The transfer fidelity in that case exhibits an increased dependence of the relative phase between   the cavity field and a squeezing device, as would be expected. In the optimal case $\theta_s =\theta_c\pm\pi/2$ and $N=10$, the fidelity of squeezed state transfer is $\mathcal{F}=0.978$, which is somewhat lower than for a vacuum state, a consequence of the higher sensitivity of squeezed states to decoherence.  


\section{Conclusion}

In summary, we have analyzed the optically mediated interaction between multimode phonon fields in cavity optomechanical systems where it is possible to adiabatically eliminate the electromagnetic field. A number of possible forms of this interaction were identified, including multimode beam-splitter interactions and, in the case of quadratic optomechanical coupling, a plethora of possible nonlinear effective interactions, such as analogs of second-harmonic generation and four-wave mixing in nonlinear optics. The specific forms of the effective interaction depend sensitively on the sign of the couplings. Importantly, we found that incoherent effects due to quantum fluctuations -- associated primarily with the elimination of the electromagnetic field -- need not overwhelm coherent mode-coupling effects. As a result one can expect that  analogs of nonlinear effects such as four-wave mixing, soliton generation, the creation of entangled phonon pairs, and others should be achievable in quantum acoustics. 

Our treatment was kept sufficiently generic that it can be applied to a wide variety of systems and parameter regimes, and as a concrete example we have discussed quantum state transfer between a momentum side-mode of a Bose-Einstein condensate and the center-of-mass mode of oscillation of the end mirror of a Fabry-P\'erot interferometer. We showed how decoherence effects can be reduced significantly in that case by the use of squeezed optical fields.  

Future work will expand these results to more general situations and apply it to the study of practical situations where the effective mode coupling is nonlinear and investigate particular effects such as phase conjugation, entanglement, and the creation of entangled phonon pairs, for example. We will also exploit pulsed optomechanics ideas to further expand the toolbox of quantum optomechanics. 

\acknowledgements

We thank E. M. Wright for stimulating discussions. This work is supported in part by the National Science Foundation, ARO, and the DARPA ORCHID and QuASAR programs through grants from AFOSR and ARO.

\end{document}